\documentclass[fleqn,twoside]{article}
\usepackage{espcrc2}

\usepackage{graphicx}
\usepackage[figuresright]{rotating}
\usepackage{epsfig}

\newcommand{\AmS}{{\protect\the\textfont2  A\kern-.1667em\lower.5ex\hbox{M}\kern-.125emS}}
\newcommand{\beqa}{\begin{eqnarray}}
\newcommand{\eeqa}{\end{eqnarray}}
\newcommand{\beq}{\begin{equation}}
\newcommand{\eeq}{\end{equation}}
\newcommand{\pslash}{p\!\!\!/\,}

\title{
{\noindent\small 
IPPP/05/67  \hspace*{1cm} DCPT/05/134}\\
Studying unquenching effects in QCD with Dyson-Schwinger equations
\thanks{Invited talk given by C.~S.~F. at the 
'Workshop on computational hadron physics', 
Sept. 13 - 17, Nikosia, Cyprus.}
}

\author{Christian~S.~Fischer\address{IPPP, University of Durham, Durham DH1 3LE, U.K.},{}
        Reinhard~Alkofer\address{Institute of Physics, Graz University,
          Universit\"atsplatz 5, A-8010 Graz, Austria},{}        
        Wolfgang~Cassing\address{Institute for Theoretical Physics, 
	  Univ. of Giessen, Heinrich-Buff-Ring 16, 35392 Giessen, Germany},{}
        Felipe~Llanes-Estrada\address{Fisica Teorica I, 
	  Univ. Complutense, Madrid 28040, Spain},{}
        Peter~Watson\address{Institute for Theoretical Physics, University of
          T\"ubingen, D-72076 T\"ubingen, Germany},{}}
       
\begin{document}

\begin{abstract}
We summarise recent results on the properties of gluons, quarks and 
light mesons from the Green's functions approach to QCD.
We discuss a self-consistent, infrared power law solution for the
Schwinger-Dyson equations of the 1PI-Greens functions of Yang-Mills 
theory. The corresponding running coupling has a universal fixed 
point at zero momentum. Based on these analytical results a truncation
scheme for the coupled system of Schwinger-Dyson equations for the 
propagators of QCD and the Bethe-Salpeter equation for light mesons 
has been formulated. We compare numerical results for charge 
eigenstate vector and pseudoscalar meson observables with corresponding 
lattice data. The effects of unquenching the system are found to be
small but not negligible. 

\vspace{1pc}
\end{abstract}

\maketitle

One of the most fascinating problems of QCD is to find a low energy 
description of colourless bound states (hadrons) in terms of their 
nonperturbative, coloured constituents (quarks and gluons). 
Lattice simulations are not entirely satisfactory in this respect. 
They provide values for the global properties of hadrons (masses, decay 
widths etc.,), but they may not be capable to provide enough information 
on their internal structure to understand all dynamical aspects of low 
energy QCD. An alternative field theoretical and relativistic method 
which is well suited to deliver this information is the Schwinger-Dyson 
and Bethe-Salpeter formalism \cite{Alkofer:2000wg,Maris:2003vk}. In 
principle this approach allows one to derive meson properties directly 
from the fundamental building blocks of the field theory, the Green's 
functions. Lattice simulations and the Green's functions approach are 
complementary to each other in several 
respects. Lattice simulations are ab initio whereas the Green's functions 
approach has to rely on an (educated) approximation scheme. 
The Green's functions approach is continuum based. 
It allows for analytical investigations in the infrared and all aspects 
of chiral symmetry and its breaking pattern are respected. In this talk 
we summarise recent results in the Green's functions framework that dwell 
on these advantages. We discuss analytical results on the infrared exponents 
of the 1PI-Green's functions of SU($N_c$)-Yang-Mills theory in Landau gauge. 
We report on numerical results for the ghost, gluon and quark propagators 
as well as light meson observables in a truncation scheme that is based on
the analytical findings. Our focus in particular is on unquenching effects 
due to light quark loops in the gluon polarisation.

\section{Infrared exponents for the Green's functions of Yang-Mills theory}

The infrared behaviour of the Green's functions of SU($N_c$)-Yang-Mills theory
is related to confinement in several ways. A particularly interesting
example is the Kugo-Ojima confinement criterion of a well-defined
global colour charge. The criterion is satisfied in Landau gauge if the dressing 
function of the ghost propagator is singular and the gluon propagator is finite 
or vanishes in the infrared. Provided BRST-symmetry is conserved nonperturbatively, 
the cohomology of the BRST-operator then defines a physical state space with 
colourless states only \cite{Nakanishi:qmas}.

\begin{figure}[t]
\centerline{\epsfig{file=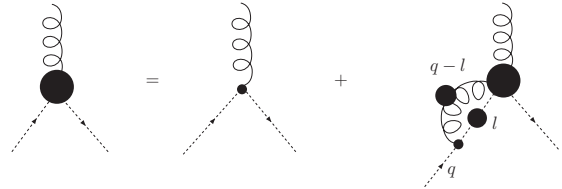,width=\columnwidth}}
\caption{Schwinger-Dyson equation for the ghost-gluon vertex.}
\label{DSE-ghg}
\end{figure}

A convenient starting point to investigate the infrared behaviour of 
one-particle-irreducible (1PI) Green's functions is the Schwinger-Dyson equation 
(SDE) for the ghost-gluon vertex, shown diagrammatically fig.~\ref{DSE-ghg}.  
The dressed ghost-gluon vertex $\Gamma_\mu^{abc} = \Gamma_\mu(p, q) f^{abc}$ 
can be represented by the two tensor structures
\beq
\Gamma_\mu(p, q) = p_\mu A(p^2,q^2) + k_\mu B(p^2,q^2),
\eeq
where $p_\mu$ and $q_\mu$ are the momenta of the outgoing and incoming
ghost and $k_\mu=-p_\mu-q_\mu$ is the gluon momentum. In Landau gauge, the 
momentum $q_\mu$ of the incoming ghost factorises from the vertex dressing, as can be 
seen from fig.~\ref{DSE-ghg}: Since the gluon propagator $D_{\mu \nu}$ is 
transverse in Landau gauge, its contraction with the bare ghost-gluon vertex 
$l_\mu$ in the loop of the SDE gives 
$l_\mu D_{\mu \nu}(l-q) = q_\mu D_{\mu \nu}(l-q)$. 

Let us assume for the moment that\\
\centerline{{\bf (I)} the loop-integral is finite in the infrared.}\\ 
(We come back to this assumption in the paragraph below eq.~(\ref{IRsolution}).)
We then observe that the dressing of the full ghost-gluon vertex vanishes 
if $q_\mu$ goes to zero,
\beq
\Gamma_\mu(p, q) = p_\mu (A-B) - q_\mu B \ \
\stackrel{q_\mu \rightarrow 0}{\longrightarrow} \ \ p_\mu,
\eeq
and thus neither $A(p^2,q^2)$ nor $B(p^2,q^2)$ can be singular in this 
limit \cite{Watson:2001yv}.
Since factorisation of the other ghost momentum is trivial, the same happens
for $p_\mu \rightarrow 0$. One thus concludes that the dressing
of the ghost-gluon vertex is finite in the infrared and may be well
approximated at small momenta by the bare vertex. This has been confirmed
by lattice and SDE-studies \cite{Cucchieri:2004sq,Sternbeck:2005qj,Schleifenbaum:2004id}.

\begin{figure}[t]
\centerline{\epsfig{file=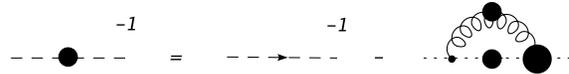,width=\columnwidth}}
\caption{Schwinger-Dyson equation for the ghost propagator.}
\label{DSE-gh}
\end{figure}

A finite ghost-gluon vertex at small momenta admits the following power law
solution for the ghost-SDE in the infrared: Writing the ghost and gluon 
propagators as
\beqa
D^G(p^2) &=& - \frac{G(p^2)}{p^2} \, , \nonumber\\
D_{\mu \nu}(p^2)  &=& \left(\delta_{\mu \nu} -\frac{p_\mu 
p_\nu}{p^2}\right) \frac{Z(p^2)}{p^2} \, ,
\eeqa
one finds power laws for the ghost and gluon dressing functions with
interrelated exponents given by
\beq
G(p^2) \sim (p^2)^{-\kappa}, \hspace*{1cm} Z(p^2) \sim (p^2)^{2\kappa}\,.
\label{kappa}
\eeq
[NB: This can be checked easily by just counting anomalous dimensions on both
sides of the equation. Note that the loop-integral is dominated by 
momenta of the same magnitude as the external momentum. Thus, for small 
external momenta one can replace the propagators in the loop by their
infrared approximation, eq.~(\ref{kappa}).] 
In this notation the Kugo-Ojima criterion translates to the condition
$\kappa \ge 0$ for the ghost dressing function and $\kappa \ge 0.5$ for
the gluon dressing function. On general grounds, the exponent $\kappa$ 
is known to be positive \cite{Watson:2001yv}, independent of any 
truncation of the SDEs. For a bare ghost-gluon vertex in the infrared
one obtains $\kappa = (93 - \sqrt{1201})/98 \approx 0.595$ 
\cite{Zwanziger:2001kw,Lerche:2002ep}, which satisfies both criteria. 
This specific value of $\kappa$ is found to vary only slightly for a large 
class of possible dressings of the ghost-gluon-vertex \cite{Lerche:2002ep}.
Similar values have been determined 
from exact renormalisation group equations \cite{Pawlowski:2003hq}.

\begin{figure}[t]
\centerline{\epsfig{file=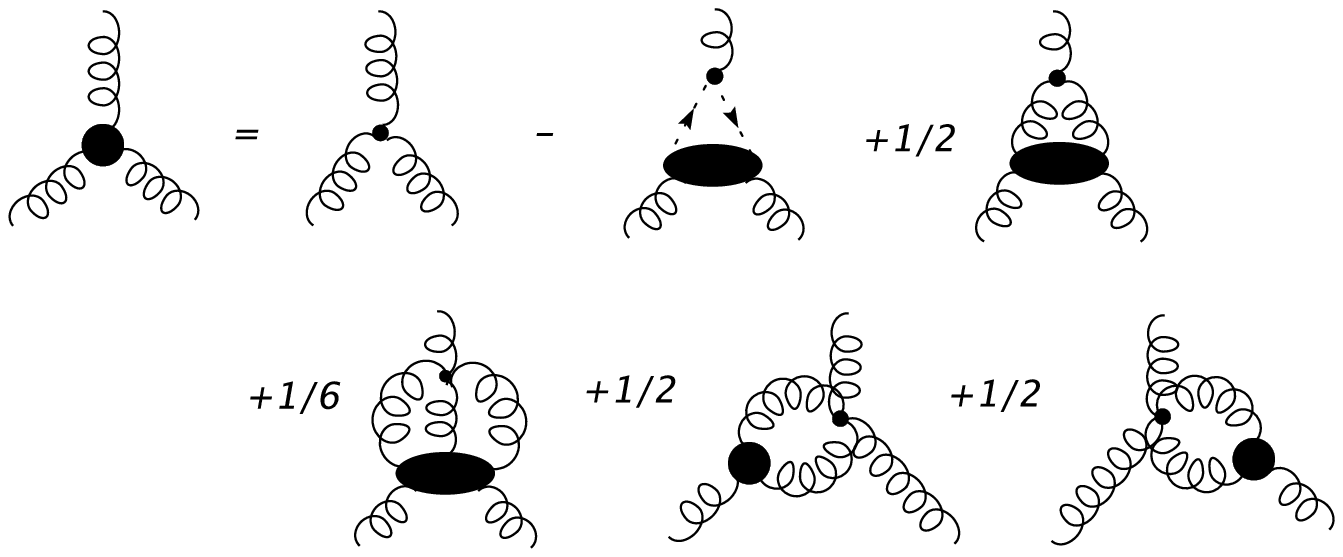,width=\columnwidth}}
\vspace*{8mm}
\centerline{\epsfig{file=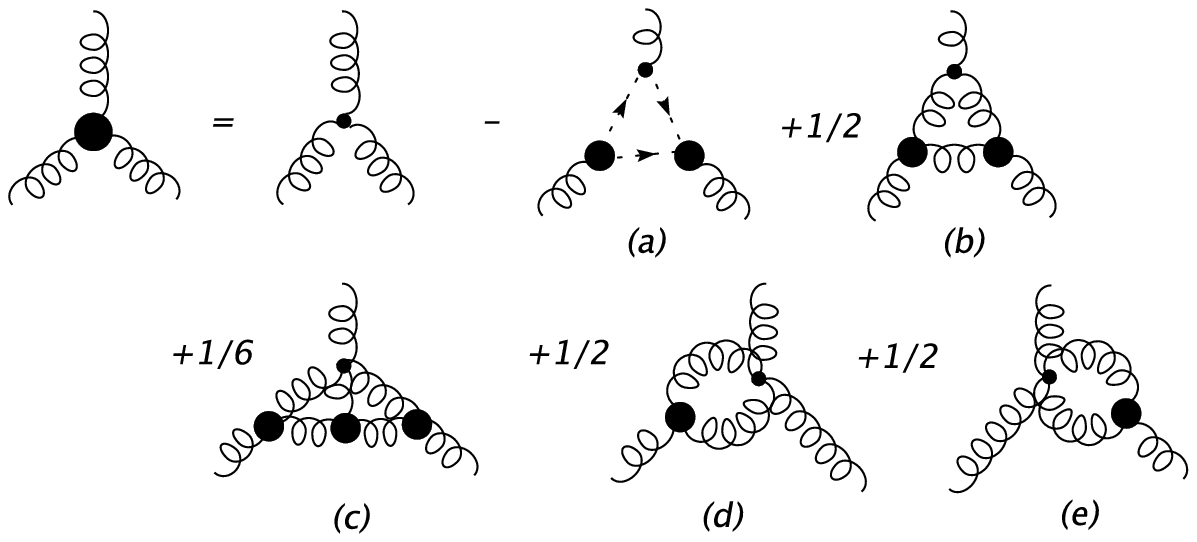,width=\columnwidth}}
\caption{Exact Schwinger-Dyson equation for the three-gluon vertex
and  lowest order in a skeleton expansion of the four-
and five-point functions. All internal propagators in the diagrams are to
be understood as fully dressed.} 
\label{DSE-3g}
\end{figure}
 
The basic idea to determine the infrared behaviour of the other
(1PI) Green's functions is to investigate their Schwinger-Dyson equations 
order by order in a skeleton expansion ({\it i.e.} a loop expansion using 
full propagators and vertices). This program has been carried out in 
ref.~\cite{Alkofer:2004it}. It turns out that in this expansion the 
Green's functions can only be infrared singular, if all external scales 
go to zero. Thus to determine the degree of possible singularities it is 
sufficient to investigate the SDEs in the presence of only one external 
scale $p^2 \ll \Lambda^2_{QCD}$. As an example we summarise the treatment 
of the SDE for the three-gluon vertex. In fig.~\ref{DSE-3g} we show the
full equation as well as the lowest order in a skeleton expansion of the 
four-and five-point functions. In the presence of one (small) external 
scale the approximated SDE has a selfconsistent power law solution given by 
\beq
\Gamma^{3g}(p^2) \sim (p^2)^{-3\kappa}. \label{IR_3g}
\eeq
[NB: Again this can be seen easily by counting anomalous dimensions on both
sides of the equations. The leading diagram on the right hand side is the
one involving ghosts, diagram (a), the others are less singular (recall $\kappa > 0$). 
The loops are again dominated by momenta of the same magnitude as the 
external scale.] One can see by induction that this solution is also 
present if terms to arbitrary high order in the skeleton expansion are taken 
into account. Thus the skeleton expansion is stable wrt. the infrared solution
of the SDEs. This technique can also be applied to any other SDE. A 
self-consistent solution of the whole tower of SDEs is then 
given by \cite{Alkofer:2004it}
\beq
\Gamma^{n,m}(p^2) \sim (p^2)^{(n-m)\kappa}. \label{IRsolution}
\eeq
Here $\Gamma^{n,m}(p^2)$ denotes the infrared leading dressing function of 
the 1PI-Green's function with $2n$ external ghost legs and $m$ external 
gluon legs. By counting anomalous dimensions it can be checked easily that
the expression eq.~(\ref{IRsolution}) indeed solves the full three-gluon vertex
SDE in fig.~\ref{DSE-3g} selfconsistently. Furthermore, inserting 
$\Gamma^{1,2}(p^2) \sim (p^2)^{-\kappa}$ together with the power laws 
(\ref{kappa}) into the SDE for the ghost-gluon vertex, fig.~\ref{DSE-ghg},
one can verify the assumption {\bf (I)} that the loop-integral of 
the vertex dressing is indeed finite in the infrared. Thus eq.~(\ref{IRsolution})
is a truly selfconsistent infrared solution of the tower of SDEs.
[NB: It is worth mentioning that the solution (\ref{IRsolution}) also has the
correct scaling behaviour such that the Slavnov-Taylor identities of
the renormalisation constants are satisfied. Since the theory is 
multiplicative renormalisable these functions scale with the 
renormalisation point $\mu^2$ in the same way as the 1PI-functions with the
external scale $p^2$. E.g. the relation $Z_1/Z_3 = \widetilde{Z}_1/\widetilde{Z}_3$
between the three-gluon vertex, gluon propagator, ghost-gluon vertex and ghost 
propagator renormalisation constant leads to $Z_1(\mu^2) = (\mu^2)^{-3\kappa}$, 
which agrees with eq.~(\ref{IR_3g}).]

Certainly, selfconsistency is not enough to establish eq.~(\ref{IRsolution})
as the 'true' solution of Yang-Mills theory in the infrared, since there may be
other selfconsistent solutions of the SDEs. However, one may
argue that the solution (\ref{IRsolution}) has an interesting property
that qualifies it as a promising candidate: it leads to 
qualitative universality of the running coupling in the infrared. 
Renormalisation group invariant
couplings can be defined from either of the primitively divergent vertices 
of Yang-Mills-theory, {\it i.e.} from the ghost-gluon vertex ($gh$), 
the three-gluon vertex ($3g$) or the four-gluon vertex ($4g$) via
\beqa
\alpha^{gh}(p^2) &=& \frac{g^2}{4 \pi} \, G^2(p^2) \, Z(p^2) 
     \hspace*{7mm} \stackrel{p^2 \rightarrow 0}{\sim} \hspace*{0mm} 
     \frac{c_1}{N_c} \,, \label{gh-gl}\\
\alpha^{3g}(p^2) &=& \frac{g^2}{4 \pi} \, [\Gamma^{0,3}(p^2)]^2 \, Z^3(p^2) 
    \hspace*{0mm} \stackrel{p^2 \rightarrow 0}{\sim}
     \hspace*{0mm} \frac{c_2}{N_c} \,,\\
\alpha^{4g}(p^2) &=& \frac{g^2}{4 \pi} \, [\Gamma^{0,4}(p^2)]^2 \, Z^4(p^2) 
    \hspace*{0mm} 
    \stackrel{p^2 \rightarrow 0}{\sim} \hspace*{0mm} \frac{c_3}{N_c} \,.
     \label{alpha}
\eeqa
Using the SDE-solution (\ref{IRsolution}) it is easy to see that all three 
couplings approach a fixed point in the infrared. The constants $c_i$ may be 
different for each coupling and depends on the respective choice of the 
tensor component used to extract the vertex dressing functions $\Gamma$ (this
ambiguity is well know in the literature \cite{Pascual:1980yu}).
For the coupling (\ref{alpha}) of the ghost-gluon vertex this fixed point can be
explicitly calculated using propagator dressing functions alone. Employing a bare
ghost-gluon vertex one obtains $\alpha^{gh}(0) \approx 8.92/N_c$
\cite{Lerche:2002ep}. Recently it has been shown that this value together with  
the infrared exponent $\kappa \approx 0.595$ are invariant
in a class of transverse gauges that interpolate between Landau and 
Coulomb gauge \cite{Fischer:2005qe}. 

\section{Unquenching effects in propagators}

\begin{figure}[t]
\vspace{0.5cm}
\centerline{
\epsfig{file=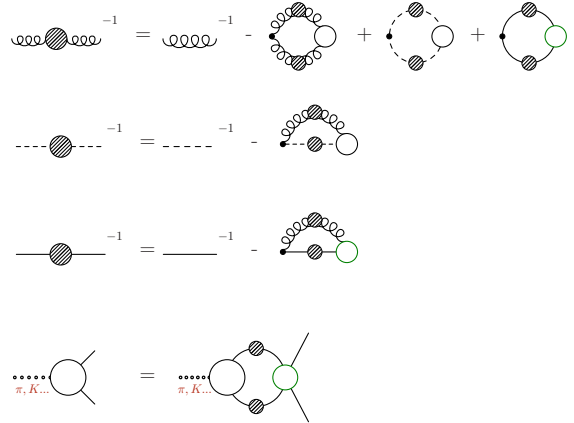,width=\columnwidth}
}
\caption{\label{SDEs} A diagrammatical representation of the coupled system of 
ghost, gluon and quark Schwinger-Dyson equations and the meson Bethe-Salpeter
equation.  Filled blobs denote dressed 
propagators and empty circles denote dressed vertex functions.
}
\end{figure}

Dynamical chiral symmetry breaking is, besides confinement, the most  
important low energy property of QCD. It is a truly nonperturbative effect,
since there is no dynamical mass generation at every order in perturbation 
theory. On the quark level, the Schwinger-Dyson formalism is well suited to
investigate the chiral symmetry breaking pattern also in the chiral limit of 
vanishing current quark masses. It is therefore complementary to lattice 
simulations, which provide reliable results for large quark masses and
volumes, but are yet severely restricted close to the chiral limit.
In the SDE-framework the effects of dynamical chiral symmetry breaking in 
quenched and (partially) unquenched QCD have been investigated in 
refs.~\cite{Fischer:2005,Fischer:2003rp}. Based on the analytical results
summarised in the last section, {\it ans\"atze} for the vertices 
have been constructed such that the system of SDEs for the ghost, gluon
and quark propagators are closed and can be solved numerically (for details
of the truncation scheme see refs.~\cite{Fischer:2005,Fischer:2002hn}). 
Here we focus in particular on unquenching effects in these propagators,
which are generated by quarks in the gluon SDE, cf. fig.~\ref{SDEs}.

\begin{figure}[t!]
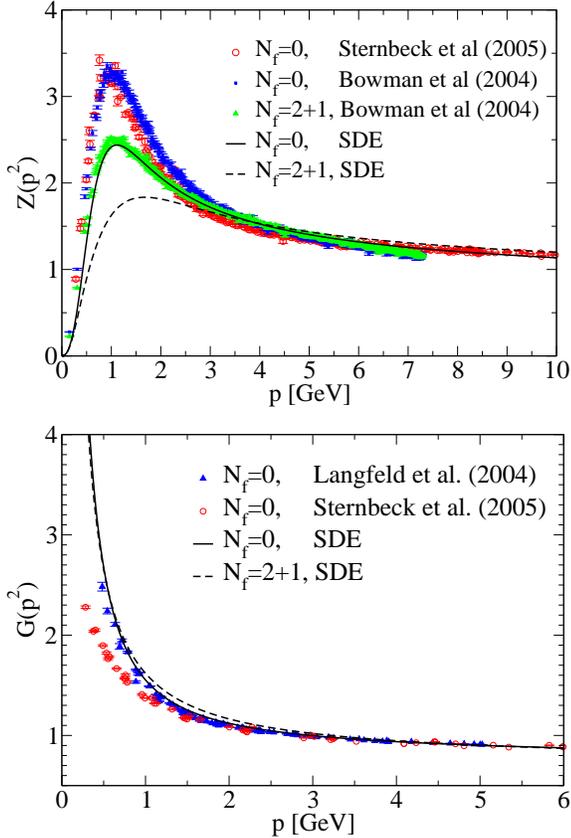

\centerline{
\epsfig{figure=GGA.gluelatt.eps,width=\columnwidth}}
\vspace*{2mm}
\centerline{
\epsfig{figure=GGA.ghostlatt.eps,width=\columnwidth}}
\caption{\label{fig:YM-dress} Comparison of the quenched and unquenched 
ghost and gluon dressing functions with recent lattice data 
\cite{Sternbeck:2005tk,Bowman:2004jm,Gattnar:2004bf}. The sea-quark 
masses are $m_{u/d} \simeq 16 \,\mbox{MeV}, m_s \simeq 79 \,\mbox{MeV}$ in 
the lattice simulations and 
$m_{u/d} \simeq 3.9 \,\mbox{MeV}, m_s \,\simeq 84 \mbox{MeV}$ in the 
SDE-approach.}
\end{figure}

Numerical solutions for the ghost and gluon propagators can be seen in 
fig.~\ref{fig:YM-dress}. In the infrared, the numerical SDE-results reproduce 
the analytical power laws, eqs.(\ref{kappa}).  (This can be seen explicitly on a 
log-log-plot, displayed e.g. in Ref.~\cite{Fischer:2002hn}.) In the ultraviolet 
they reproduce the correct one-loop running from resummed perturbation theory. 
Compared to the results of recent lattice calculations 
\cite{Sternbeck:2005tk,Bowman:2004jm,Gattnar:2004bf} (see also \cite{Oliveira:2004gy})
we find good agreement for large and small momenta. Small deviations for the 
value of the infrared exponent $\kappa$ between continuum-SDE and lattice
results on a finite volume may have methodical reasons. The resulting 
running coupling on the lattice does not reproduce the fixed point behaviour
in the continuum but vanishes in the infrared 
\cite{Sternbeck:2005tk,Alles:1996ka,Boucaud:2005qf}. These effects are also seen 
when one solves Schwinger-Dyson equations on a torus and are discussed in detail 
elsewhere \cite{Fischer:2005ui} (see also \cite{Pawlowski:2004ip}). In the intermediate 
momentum region one clearly sees unquenching effects in the gluon dressing function 
\begin{figure}[t!]
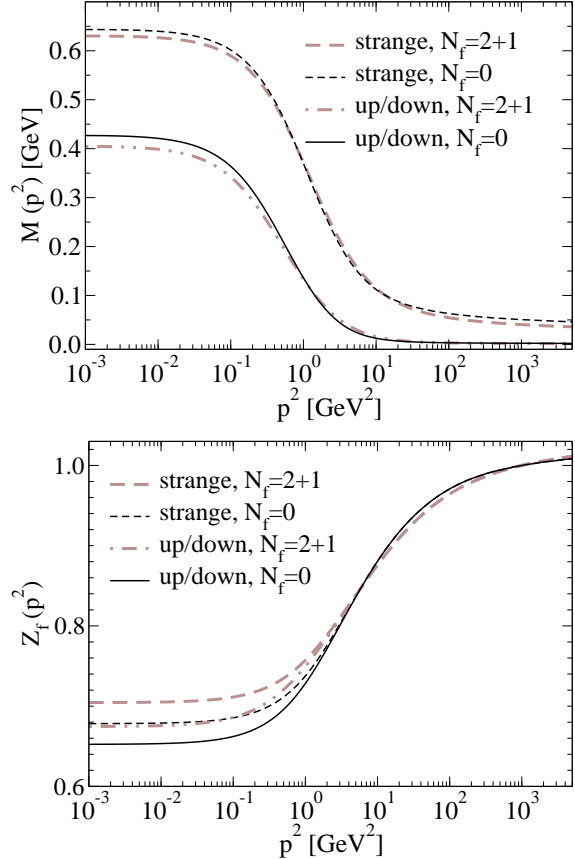

\centerline{
\epsfig{file=GGA.M.lin2.eps,width=\columnwidth}}
\vspace*{2mm}
\centerline{
\epsfig{file=GGA.A2.eps,width=\columnwidth}}
\caption{\label{fig:qspace_full} Comparison of the quenched and unquenched 
quark mass (upper diagram) and wave-functions (lower diagram).}
\end{figure}
due to the formation of quark-antiquark pairs from the vacuum. The screening effect 
from these pairs decreases the bump in the gluon dressing function considerably.
The overall difference in the size of the bump between the lattice and the SDE-results
is a measure of the influence of the (neglected) gluon-two-loop diagrams in the 
gluon-SDE.

The up/down and strange quark propagator functions,
\beq
S(p) = Z_f(p^2)/(-i \pslash + M(p^2))
\eeq
 are plotted 
in fig.~\ref{fig:qspace_full}.  We clearly observe a large amount of dynamical mass 
generated in the infrared. This mass is reduced by roughly 10 percent once 
quark-loops are taken into account. For large momenta the numerical solutions 
reproduce the logarithmic running known from resummed perturbation theory. 
There are noticeable unquenching effects in the intermediate momentum region, which
are, however, much smaller in size than those observed in the gluon propagator.
By explicitly solving the quark-SDE in the complex plane one finds a pair of complex
conjugate poles at $(0.47\pm0.29\imath) \mathrm{GeV}$ (quenched) 
and $(0.45\pm0.27\imath) \mathrm{GeV}$ (unquenched). Thus unquenching hardly has any
effect on the position of these singularities (see however ref.~\cite{Alkofer:2003jj} 
for a discussion of the possible influence of scalar tensor pieces in the quark-gluon 
vertex on the analytical structure of the propagator).

\section{Unquenching effects in light meson observables}

\begin{figure}[t!]
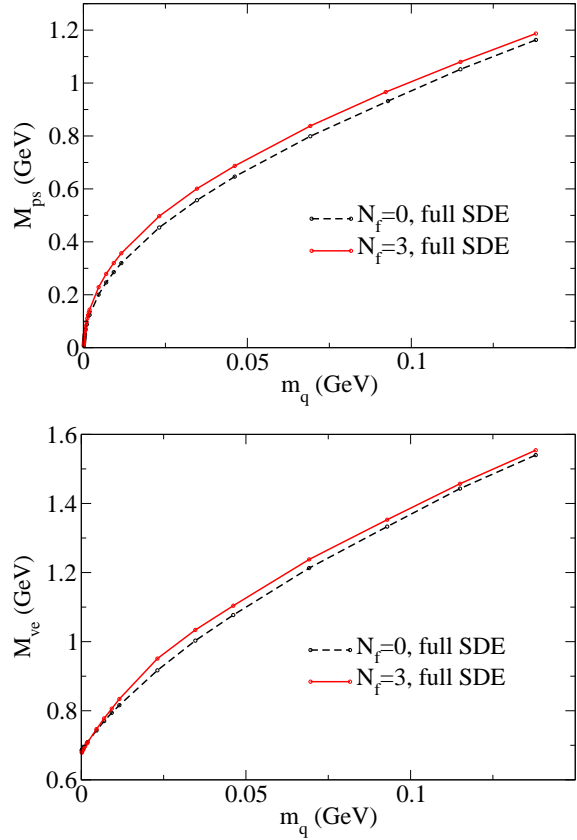

\centerline{
\epsfig{file=pi_full.eps,width=\columnwidth}}
\vspace*{2mm}
\centerline{
\epsfig{file=rho_full.eps,width=\columnwidth}}
\caption{\label{fig:plot1} Pseudoscalar (upper diagram) and vector (lower diagram) 
meson masses as functions of the quark mass parameter.  We compare results for 
the quenched and unquenched theory using three degenerate sea quarks.}
\end{figure}

The results for the gluon, ghost and quark propagators, discussed in the last section,
serve as input for a calculation of light meson observables employing a
Bethe-Salpeter equation (BSE), cf. fig.~\ref{SDEs}. The crucial link between the bound 
states and their quark and gluon constituents is provided by the axialvector 
Ward-Takahashi identity. It relates the quark self energy with the 
quark-quark interaction kernel in the BSE and thereby guarantees the Goldstone
nature of the pions and kaons \cite{Maris:1997hd}. For details of the implementation
of this identity within the truncation scheme discussed here see ref.~\cite{Fischer:2005}.
Results for the masses of the pseudoscalar and vector meson (in the isospin symmetric limit) 
as a function of the current quark mass are shown in fig.~\ref{fig:plot1}. 
For $m_q\rightarrow 0$ (the chiral limit) we observe a massless pion as expected.
As for unquenching effects both the pseudoscalar and vector masses 
with larger quark masses are increased $\sim 30\mathrm{MeV}$ when quarks loops
are taken into account.

The values for the current quark masses at the physical point together with the 
resulting pseudoscalar meson masses, leptonic decay constants and vector meson 
masses are given in Table~\ref{tab:fpar}. When fitted to the experimental pion 
and kaon masses the resulting up/down and strange-quark masses are lowered when 
quark loop effects are taken into account. This has also been observed in 
corresponding lattice simulations \cite{CP-PACS,JLQCD}. We furthermore see that 
the results for $f_K$ and $m_{\rho}$ are quite insensitive to whether or not the 
system is unquenched (in the restricted 
sense of fig. 4).  This leads to the conclusion that once the 
interaction has been fitted to the pseudoscalar observables, the vector 
\begin{table}[t]
\begin{center}\begin{tabular}{|c|c|c|c||c|}\hline
$N_f$    &  0      & 3           &  2+1        & PDG \cite{Eidelman:2004wy} \\\hline\hline
$m_u$    & 4.17    & 4.06        & 4.06        & 3-5 \\\hline
$m_s$    & 88.2    &             & 86.0        & 80-130  \\\hline
$m_\pi$  & 139.7   & 139.7       & 140.0       & 139.6  \\\hline
$f_\pi$  & 130.9   & 131.1       & 131.0       & 130.7  \\\hline
$(-\langle \bar{q}q\rangle)^{1/3}$ 
         & 266     & 271         & 271         &        \\\hline
$m_K$    & 494.5   &             & 493.3       & 493.7  \\\hline
$f_K$    & 165.6   &             & 169.5       & 160.0  \\\hline
$m_\rho$ & 708.0   & 690.0       & 695.2       & 770.0  \\\hline
\end{tabular}\end{center}
\caption{\label{tab:fpar} Parameter sets and results for $m_{\pi}$, 
$f_{\pi}$, $m_K$, $f_K$ and $m_{\rho}$ for the quenched
case ($N_f=0$), the unquenched case with three degenerate 'sea'-quarks 
($N_f=3$) and the physical quark configuration case ($N_f=2+1$) with two 
up/down quarks and one strange quark. The quark masses and the condensate 
have been determined using a large renormalisation point
and subsequently evolved down to the scale
$\mu = 2\,\mathrm{GeV}$ according to their one-loop running.
All units are MeV.}
\end{table}
meson mass is largely fixed.  The likely explanation for this is that the 
ground state pseudoscalar and vector mesons are both states with the lowest 
orbital angular momentum ($L=0$) in the sense of the naive (quantum mechanical) 
quark model -- meaning that they are determined largely by the lowest spin 
contributions of the kernel in the Bethe-Salpeter equation given by the 
ladder approximation (which is used here).  The interaction then plays the 
same role in both channels, hence the similarity in results.  Note that the 
$\rho$ meson calculated within the framework of the truncated Bethe-Salpeter 
equation here refers to a pure quark-antiquark meson with no allowed decay 
channel. A first step towards including the non-trivial decay width of the
physical $\rho$ meson in this formalism has been made in 
ref.~\cite{Watson:2004jq} 
(see also ref~\cite{Jarecke:2002xd} for a calculation 
of $g_{\pi \rho \rho}$ using quenched Bethe-Salpeter amplitudes).
Here the decay of $\rho$ meson will lead
to an additional shift of the $\rho$ meson mass peak as evident from 
dispersion relations. As a consequence our present results should not
directly compared to experiment. Although the $\rho$ mass is slightly
low the dispersive corrections via its two pion decay 
(and the inclusion of quark-gluon vertex corrections) 
might yield a satisfactory answer.

\begin{figure}[t!]
\centerline{
\epsfig{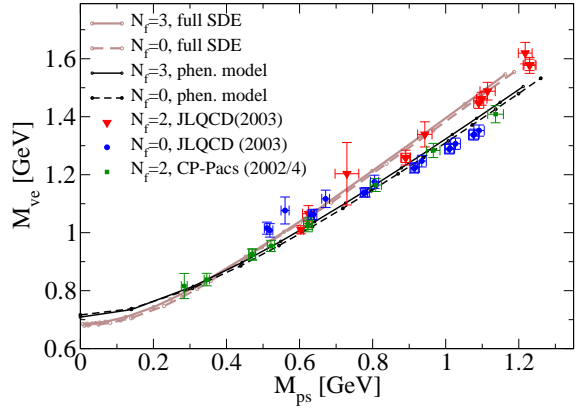}
}
\caption{\label{pive} Vector meson masses as a function of pseudoscalar 
meson masses. We compare the quenched and degenerate unquenched cases. The 
lattice results are taken from refs.~\cite{CP-PACS,JLQCD}.}
\end{figure}

A suitable quantity to compare results from the Green's functions framework
with lattice data is the vector meson mass as a function of the pseudoscalar 
meson mass. Since both quantities are physical no scheme ambiguities arise.
The results are shown in fig.~\ref{pive}. From the SDE/BSE-approach
we compare results from the full SDE-setup, described above, with those 
obtained employing a phenomenological model for the quark-gluon interaction (the 
details of the model are described in ref.~\cite{Fischer:2005}). 
The model interaction and the effective interaction of 
the full SDE-setup are complementary to each other in the sense that the
model interaction is confined to a quite narrow momentum region, whereas
the interaction of the full SDE-setup has considerable strength in the infrared
and extends into the ultraviolet according to the correct one-loop
scaling known from perturbation theory. Together, both setups represent
a measure for the theoretical error of our calculation. This error is obviously
of the same size as the combined systematic error of the different lattice 
simulations. In general, the results are in nice agreement with the lattice data.
For pion masses below 240 MeV, where no lattice data are available, the results
show a nonlinear dependence of the vector meson mass on the pseudoscalar one.
The effect of unquenching -- when viewed as a function of the pseudoscalar 
meson mass -- becomes the same for both schemes: the vector meson mass is 
slightly increased when quark loops are taken into account. This trend is 
also seen in the lattice simulations \cite{CP-PACS,JLQCD}, where the effect 
is even more pronounced.  However, these unquenching effects are small 
compared to the differences between both, the truncation schemes we employed 
and the systematic errors of the lattice results. 

{\bf Acknowledgements}

It is a pleasure to thank the organisers of the {\it Workshop on computational
hadron physics} for all their efforts which made this highly interesting conference 
possible. C.~S.~Fischer and P.~Watson thank M.~Pennington for inspiring discussions.  
The work summarised here has been supported by a grant from the Ministry of Science, 
Research and the Arts of Baden-W\"urttemberg (Az: 24-7532.23-19-18/1 and 
24-7532.23-19-18/2), the Deutsche Forschungsgemeinschaft (DFG) under contract 
Fi 970/2-1, the Virtual Institute for Dense Hadronic Matter and QCD Phase Transitions
and the Spanish FPA2005-02327.

\end{document}